\documentclass[manuscript,screen]{acmart}

\newcommand\tab[1][1cm]{\hspace*{#1}}

\AtBeginDocument{%
	\providecommand\BibTeX{{%
			\normalfont B\kern-0.5em{\scshape i\kern-0.25em b}\kern-0.8em\TeX}}}

\setcopyright{acmcopyright}
\copyrightyear{2021}
\acmYear{2021}
\acmDOI{10.1145/1122445.1122456}

\usepackage{subcaption}

\begin{document}
	
	\title{Continuous Operator Authentication for Teleoperated Systems Using Hidden Markov Models}
	
	\author{Junjie Yan}
	\email{junjiey@uw.edu}
	\affiliation{%
		\institution{University of Washington}
		\streetaddress{18550 NE 53rd Ct}
		\city{Redmond}
		\state{WA}
		\postcode{98052}}
	
	\author{Kevin Huang}
	\email{kevin.huang@trincoll.edu}
	\affiliation{%
		\institution{Trinity College}
		\streetaddress{Engineering Department, Trinity College, 300 Summit St}
		\city{Hartford}
		\state{CT}
		\postcode{06106}}

	\author{Kyle Lindgren}
	\email{kyle509@uw.edu}
	\affiliation{%
		\institution{University of Washington}
		\streetaddress{Department of Electrical \& Computer Engineering, University of Washington,  Paul Allen Center, 185 E Stevens Way NE AE100R}
		\city{Seattle}
		\state{WA}
		\postcode{98195-0005}}
	
	\author{Tamara Bonaci}
	\email{t.bonaci@neu.edu}
	\affiliation{%
		\institution{Northeastern University}
		\streetaddress{Khoury College of Computer Sciences, Northeastern University - Seattle 401 Terry Ave N}
		\city{Seattle}
		\state{WA}
		\postcode{98109}}
	
	\author{Howard J. Chizeck}
	\email{chizeck@uw.edu}
	\affiliation{%
		\institution{University of Washington}
		\streetaddress{Department of Electrical \& Computer Engineering, University of Washington,  Paul Allen Center, 185 E Stevens Way NE AE100R}
		\city{Seattle}
		\state{WA}
		\postcode{98195-0005}}
	
	\renewcommand{\shortauthors}{Yan, et al.}
	
	\begin{abstract}
		In this paper, we present a novel approach for continuous operator authentication in teleoperated robotic processes based on Hidden Markov Models (HMM). While HMMs were originally developed and widely used in speech recognition, they have shown great performance in human motion and activity modeling. We make an analogy between human language and teleoperated robotic processes (i.e. words are analogous to a teleoperator's gestures, sentences are analogous to the entire teleoperated task or process) and implement HMMs to model the teleoperated task. To test the continuous authentication performance of the proposed method, we conducted two sets of analyses. We built a virtual reality (VR) experimental environment using a commodity VR headset (HTC Vive) and haptic feedback enabled controller (Sensable PHANToM Omni) to simulate a real teleoperated task. An experimental study with 10 subjects was then conducted. We also performed simulated continuous operator authentication by using the JHU-ISI Gesture and Skill Assessment Working Set (JIGSAWS). The performance of the model was evaluated based on the continuous (real-time) operator authentication accuracy as well as resistance to a simulated impersonation attack. The results suggest that the proposed method is able to achieve 70\% (VR experiment) and 81\% (JIGSAWS dataset) continuous classification accuracy with as short as a 1-second sample window. It is also capable of detecting an impersonation attack in real-time.
	\end{abstract}

	

	\maketitle

	\section{Introduction}
Teleoperated robotic systems have become an emerging and popular technology, largely due to several salient benefits. Critically, teleoperation provides a means to extend human capability to spaces that are otherwise inaccessible by human beings. The task may, for example, be too dangerous, such as in radioactive or chemically caustic environments, or even disaster scenarios. Furthermore, the task may be at a scale too large or too small for a human to physically accomplish. Consider the case where a human's particular expertise is not locally accessible or available, e.g. a specialized surgeon is needed on another continent. In these cases, the use of a remote robot-controlled at a distance, via a human operator, provides a practical solution. However, the benefit of having teleoperators comes with its own set of problems: \emph{what if the security of teleoperated robotic systems is compromised?} In many envisioned high-reward scenarios, basic infrastructure may be limited. Remote robots may have to operate in a harsh environment. The open and relative uncontrollable nature of the communication link between the operator and the robot potentially makes the teleoperated robotic system more vulnerable to various kinds of attacks. In our prior work \cite{bonaci2015make, bonaci2015experimental}, we discovered that the tension between real-time operation (usability) and security for teleoperated robotic systems may render many existing security techniques infeasible. Existing methods, such as data encryption and commands signature verification, potentially generate an extra burden on the communication between the operator and the remote robot, which introduces delay and thus degrades the usability of the teleoperated system. Many teleoperated robotic systems deal with delicate or critical tasks. It is, therefore, crucial to make them secure without affecting their usability. The specific solution we propose exploits the fact that there is a human operator in the loop. Human operators have unique ways of interacting with teleoperated robotic systems\cite{yan2015haptic}, and this unique operating signature can be used to identify and authenticate the operator, thus enhancing the security and privacy properties of teleoperated robotic systems. In \cite{yan2015haptic}, a new password system based on the user's behavior biometric is developed to fulfill an initial login authentication task. However, if the malicious party targets the post authenticated session, such as through communication channel jamming or `hijacking', initial authentication may not be sufficient. 

Therefore, continuous authentication needs to be developed to secure the teleoperated robotic system. Behavior biometric-based continuous authentication has emerged recently to mitigate the security problems and attacks that target the post-authenticated session after the initial login for computer systems\cite{araujo2005user, tappert2010keystroke, monaco2012developing} and mobile devices \cite{frank2013touchalytics, shi2011senguard, feng2012continuous, bo2013silentsense, roy2014hmm,sitova2015hmog,li2020scanet}. In these works, instead of authorizing a user through a one-time login challenge, the authentication system continuously examines the user's behavior biometrics (i.e. keystroke/mouse dynamics, touch screen usage, device dynamics, etc.) in order to guarantee the identity of the initially authenticated user.

In the aforementioned works, the authentication results are based on the analysis of relatively simple user actions. In \cite{araujo2005user} and \cite{tappert2010keystroke}, features of user keystroke action, such as key code, press time and interval time between strokes when a user interacts with a desktop computer, are analyzed. Touch actions on mobile devices, such as tapping, scrolling, and flinging, are used for authentication purposes. Using single actions to fulfill the authentication is highly volatile. In most cases, to increase the robustness of the authentication method, multiple consecutive actions are used for the final decision. However, the operator's motions and actions during a teleoperated procedure are far more complex than a single simple action. The methods developed in the above works are not suitable in this case. 

In \cite{alemzadeh2016targeted}, the author proposed a real-time detection method based on the teleoperated robots' dynamic properties. They focused on the case where the attackers are able to gain robot log access, analyze the teleoperation process, and inject malicious commands into the robot control system at the desired critical time. The detector estimates the robot motor, joints, and end-effector position and orientation after executing the given commands. The detector will then raise alerts whenever the velocity/acceleration exceed a pre-defined threshold. However, once the attackers gain full access to the remote robot, they will be able to cause damage without triggering the detector, such as cutting benign tissues with steady motion during a teleoperated surgery, or prematurely trigger an explosion in an unsafe location during a teleoperated bomb disposal task.

Moreover, in most of these approaches, conventional classifiers, such as k-Nearest Neighbour, Support Vector Machine, Neural Network or Random Forests, are implemented. \emph{The major limitation of these classification algorithms is that they heavily rely on the choices of both positive and negative samples during the training process.}

Although the choices of positive samples are straightforward in our case, as we can use data obtained from the genuine operator, negative samples are not so easily acquired. The relevant teleoperated task data are nonlinear, dynamic and high dimensional. Thus, creating negative samples from human-generated spoofs of teleoperated tasks would not tractably produce reliably complex sets of data that model the real-world. On the other hand, using other operators' data as negative samples for training is volatile, as classification performance will be very sensitive to the choice of these samples.

In this work, by making the analogy between human language and the motion commands of the remote operator, we present a Hidden Markov Model (HMM) based method for continuous teleoperator authentication. The HMM can be trained with only the operator's data (positive samples)\cite{roy2014hmm} and it has been widely used in speech recognition\cite{huang1990hidden} and human motion modeling \cite{sminchisescu2006conditional, shi2009towards}, which fits the teleoperator continuous authentication task well. The operator behavior-based continuous authentication is accomplished on the remote robot side with the operator's kinematic data. It gurantees that the operation performed by the remote robot is authenticated. The authentication process can be fulfilled in parallel with the teleoperation process, which minimizes its effect on the usability of the teleoperated system. 

To determine the feasibility of our approach, we performed an experimental study with 10 participants. All subjects carried out a simplified Fundamentals of Laparoscopic Surgery (FLS) block transfer task. It is a standard test used to train and test surgeons\cite{lum2008objective}. We developed a simulated virtual reality environment with haptic feedback and virtual fixtures enabled for this task. We also explored the performance of our approach on the da Vinci surgical robotic platform, as we performed a simulated continuous operator authentication task by using JHU-ISI Gesture and Skill Assessment Working Set (JIGSAWS) \cite{gaojhu}.

In summary, the main contributions of this work are:
\begin{enumerate}
	\item The development of a continuous teleoperator authentication method that uses Left-Right HMM\cite{young1997htk} to model an operator's gestures followed by a Token Passing algorithm \cite{young1989token} that concatenates gesture models.
	\item The development and demonstration of a VR simulated teleoperation environment and the experimental user study evaluation.
	\item Experimental demonstration that the proposed continuous teleoperator authentication is able to achieve high accuracy and impersonation attack resistance.
\end{enumerate}
	
	\section{Related Work}
HMMs have been extensively used in surgical skill assessment \cite{reiley2009task, reiley2008automatic, rosen2006generalized, rosen2001markov, rosen2002task, ahmidi2017dataset}. In the majority of these efforts, it is assumed that the entire surgical process is generated from a single HMM model while each surgical gesture is represented by a single state in the HMM. In \cite{rosen2002task}, it is assumed that each surgical gesture can be represented by samples from a Gaussian distribution. In \cite{reiley2009task}, the Short Time Fourier Transform (STFT) followed by K-Means is used to discretize the surgical process data into gestures (states in HMM). Discrete HMMs are then trained to fulfill the skill evaluation. Linear discriminant analysis (LDA) is applied to the surgical data to perform dimension reduction in \cite{varadarajan2009data} before HMMs are trained. In \cite{tao2012sparse}, a Sparse HMM approach is proposed, where a sparse dictionary learning technique called K-SVD\cite{aharon2006img} is used to model the surgical gesture states in the HMM. Representing a surgical gesture by a single state in HMMs has limitations with regard to fully capturing the dynamic and complex properties of each gesture. In \cite{ahmidi2017dataset, varadarajan2009data}, the authors proposed to represent each surgical gesture as an HMM instead of a single state within an HMM.

\begin{figure}
	\centering
	\includegraphics[width=0.8\columnwidth]{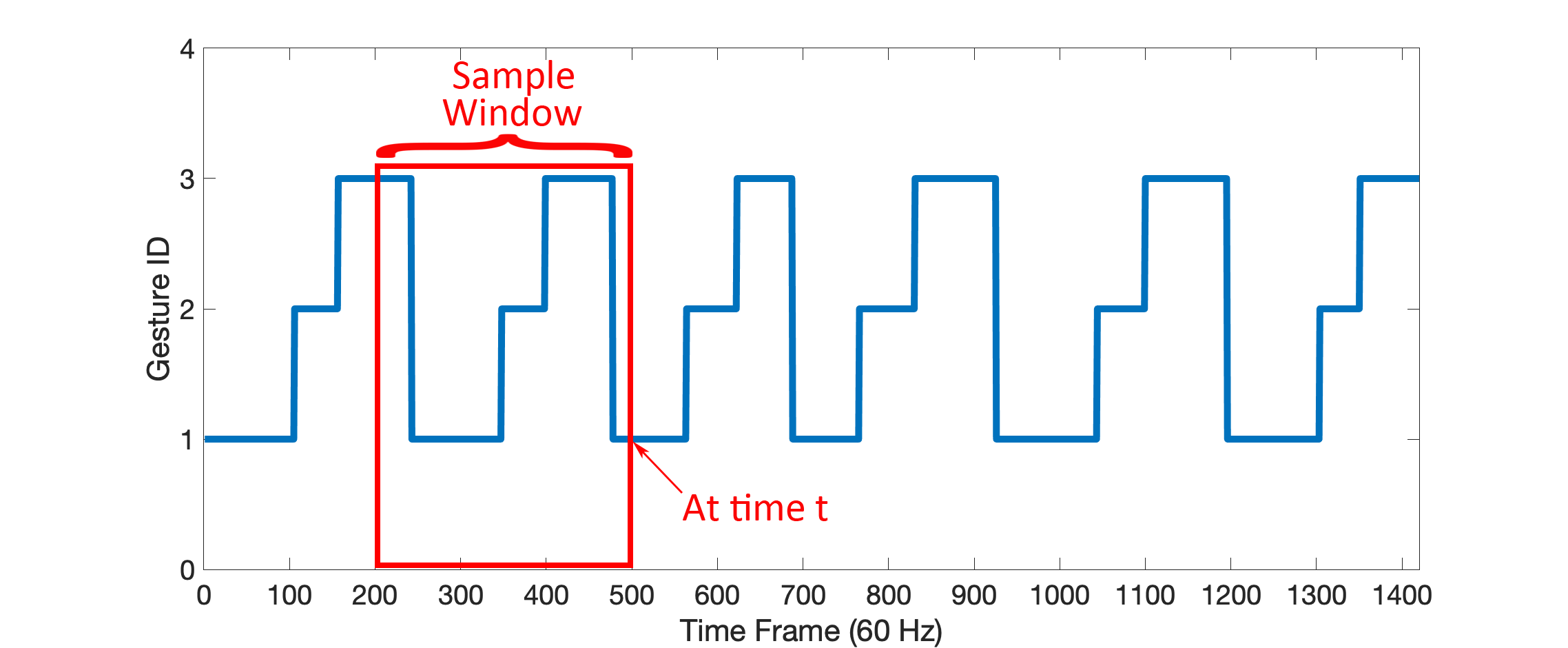}
	\caption{Comparison Between Offline Analysis and Continuous Authentication}
	\label{Fig_gest}
	
\end{figure}

In all aforementioned work, the analysis is performed offline given the entire kinematic data of the surgical procedure. However, as shown in Figure \ref{Fig_gest}, continuous authentication is an online process, and instead of the whole offline data set, we need to rely on the data from some sample window to perform analysis and authenticate an operator. In this case, unlike the offline scenario, the data in the sample window will contain only partial gestures. The kinematic properties of a partial gesture are different from the complete gesture. Therefore, using a single state in the HMM to represent the operator's gesture is not applicable for continuous operator authentication. Representing each operator's gesture as an HMM better fits our application.

In \cite{yasar2020real}, the author proposed Recurrent Neural Networks (RNN) and Long short-term memory (LSTM) networks to model surgical gestures among multiple surgeons and achieved real-time unsafe events detection in robot-assisted surgeries. In \cite{lin2006towards, tao2013surgical, lea2016temporal, lea2016segmental,lea2017temporal, ahmidi2017dataset}, automatic sugical gesture segmentation has been extensively investigated and explored. 

In this paper, our scope is to use gesture labeled teleoperation process sequences to train operators' gesture models. We then focus on using the trained operators' gesture models to continuously authenticate the corresponding operator given the unlabeled real-time teleoperation process sequences.
	\section{Threat Model and Detection Strategy}

In this paper, we focus on the detection and mitigation of {\bf impersonation attacks}. Impersonation attacks against teleoperated systems represent an advanced, persistent threat against complex teleoperated systems, especially those used in safety-critical missions. 

We assume an attacker with enough computational resources and knowledge about the system to gain access to a remote robot. The attacker can gain such an access by exploiting the vulnerable software/hardware supply channel\cite{solarwind}, by exploiting exiting software vulnerabilities \cite{falliere2011w32, lipp2018meltdown, kocher2019spectre, msexchangeserver}, by targeting vulnerable communication channels, by using stolen credentials, or through insider attacks.

Once gained access, the attackers' goal is to stay stealthy within the system, and cause damage/loss to the system users over an extended period of time. In doing so, an attacker is likely to want to impersonate a legitimate, already authenticated teleoperator, and perform operations using their credentials. 

In order to detect such an attack, we focus on the kinematic motion commands received by the remote device, presumably sent from the operator. By continuously analyzing the motion commands, the proposed detection method is able to determine whether the commands are sent from the authorized operator in real-time. It guarantees that the operation performed by the remote robot is authenticated.

\section{Continuous Operator Authentication for Teleoperated Systems}

\begin{figure*}
	\centering
	\includegraphics[width=1\columnwidth]{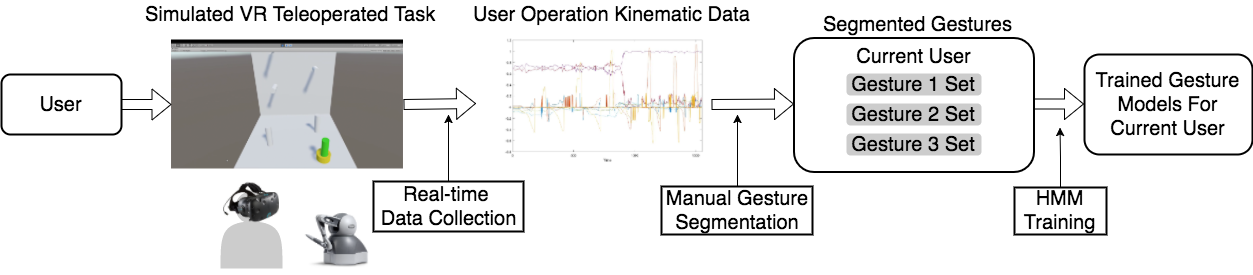}
	\caption{Continuous Operator Authentication Training Phase. The user performs a simulated teleoperation in the VR environment with a haptic input device. Real-time kinematic data of the entire teleoperation process is collected. We manually segment the teleoperation process into several basic gestures. The corresponding user's gesture HMM models are trained based on the segmented gesture pieces.}~\label{fig_cont_train}
\end{figure*}

\begin{figure*}
	\centering
	\includegraphics[width=1\columnwidth]{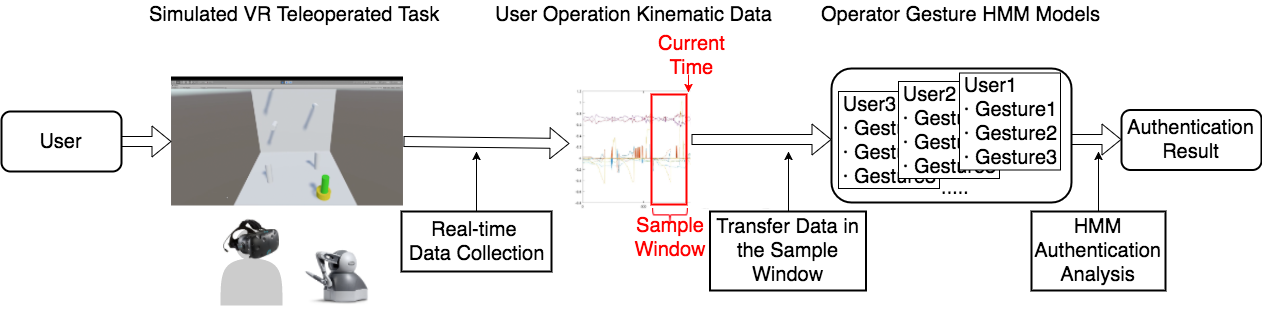}
	\caption{Continuous Operator Authentication Testing Phase. The user performs the simulated teleoperation task while we use a moving sample window to collect kinematic data. Given the  operator gesture HMM models obtained in the training phase, an HMM likelihood analysis is performed on the kinematic data within the sample window. The continuous authentication result is based on the likelihood analysis}~\label{fig_cont_test}
\end{figure*}
In order to achieve continuous operator authentication for teleoperated systems and overcome the constraints present in existing methods, we develop a novel scheme as shown in Figure \ref{fig_cont_train} (training process) and \ref{fig_cont_test} (continuous authentication process). In the training phase, a user performs a simulated teleoperation in the VR environment by using a haptic input device. Real-time kinematic data 
(i.e. velocity, orientation, and force applied) of the entire teleoperation process is collected. We then manually segment the teleoperation process into several basic gestures. The corresponding user's gesture HMM models are trained based on the segmented gesture pieces. In the testing phase, a user performs the simulated teleoperation task while we use a moving sample window with width $T$ to collect kinematic data from $t-T$ to $t$, where $t$ is the current time. An HMM likelihood analysis is then performed on the data within the sample window based on the trained operator gesture HMM models and this generates the continuous authentication result.

	\section{Experiment}
\subsection{Experimental Setup}
\begin{figure}
\begin{subfigure}{0.49\columnwidth}
	\centering
	\includegraphics[width=1\columnwidth]{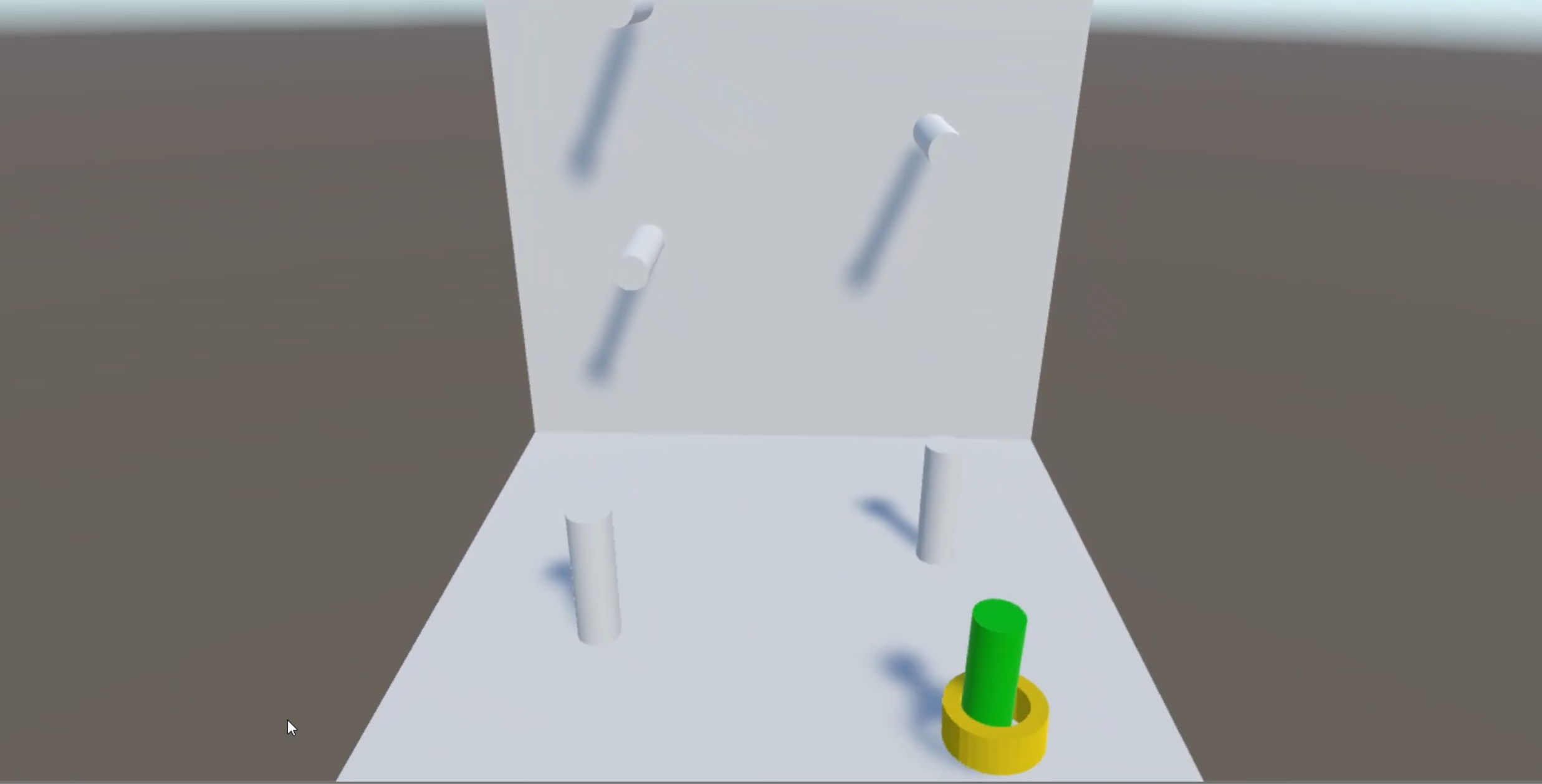}
	\caption{VR Environment}
\end{subfigure}
\begin{subfigure}{0.49\columnwidth}
	\centering
	\includegraphics[width=0.9\columnwidth]{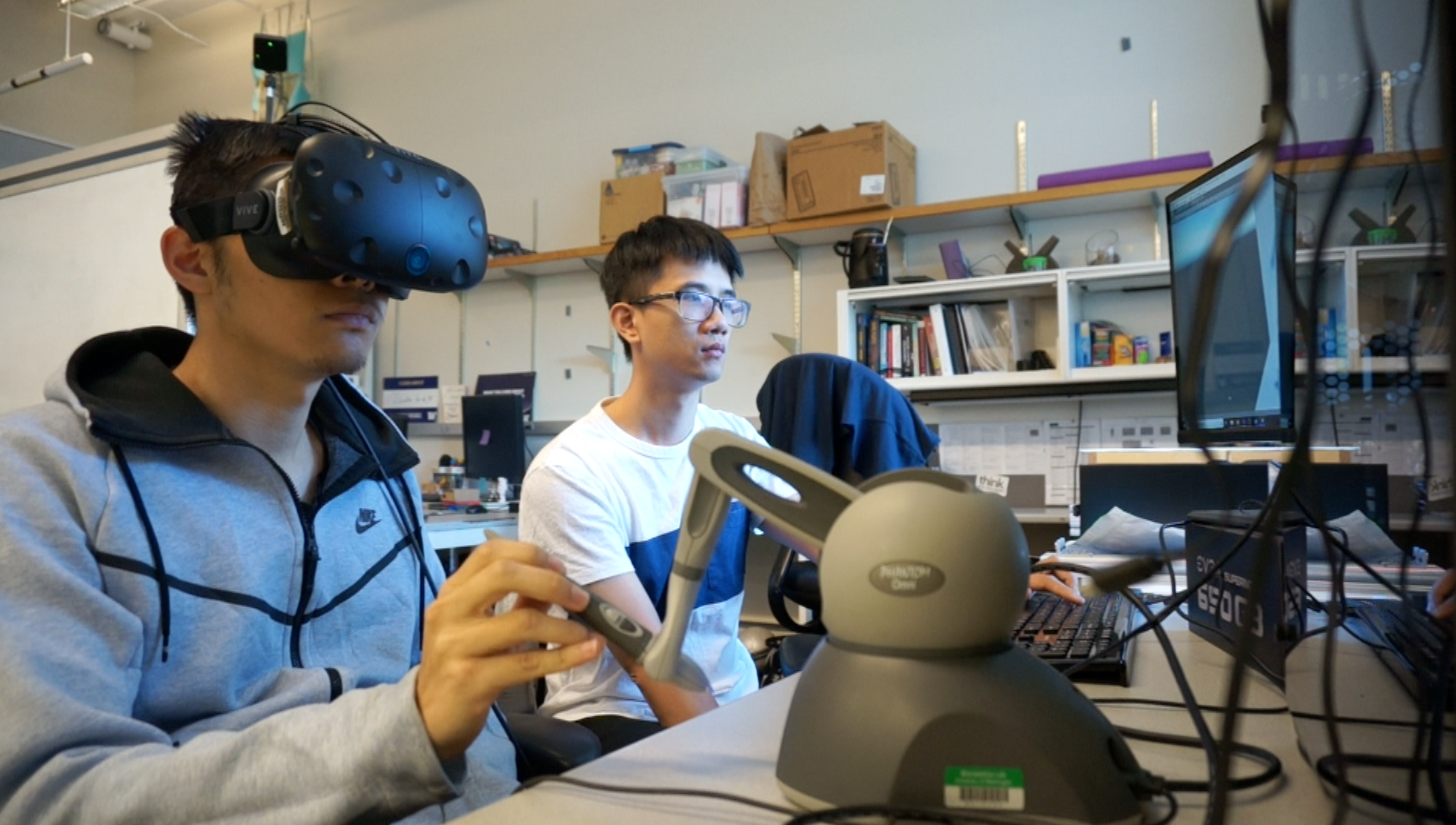}
	\caption{PhantomOmni Controller}
\end{subfigure}
	\caption{Experiment: The VR Environment (a) and The user and PhantomOmni Controller (b)}~\label{fig:cont_exp_figure1}
\end{figure}

In this work, we first built a VR environment within the Unity Game Engine\cite{engine2008unity} to let subjects perform a simplified FLS block transfer task. As shown in Fig. \ref{fig:cont_exp_figure1}, the user was asked to use the Sensable PHANToM Omni\cite{phantom2sensable} to control the 6 degree of freedom (DOF) configuration of a virtual ring in order to transfer it through the virtual pegs on the board in a predefined sequence. 

To assess the applicability of our proposed approach to real-time teleoperator authentication, we create a simplified Fundamentals of Laparoscopic Surgery (FLS) block transfer task, a standard test used to train and test surgeons\cite{lum2008objective}. The FLS block transfer task is a well-established surgical task, which is simple enough for surgical non-experts to execute, yet complex enough to segment into meaningful surgical gestures.

In the experiment, each user wore the HTC Vive headset for 3D visual feedback from the VR simulated task. Meanwhile, haptic feedback via virtual fixtures\cite{rosenberg1993virtual} was enabled during the entire operation as the user-provided motion commands with the Sensable PHANToM Omni. Two types of virtual fixtures were introduced: 1) Forbidden region around the pegs and baseboard and 2) Guidance toward the next peg tip. The guidance virtual fixture was only activated when the ring was out of the peg and being transferred toward the next peg. The haptic feedback offers the user a sense of touch and helps improve the operational performance. Moreover, in \cite{yan2015haptic}, we found that humans have unique ways of interacting with haptic interfaces and that haptic feedback can be used for continuous authentication.

In this experiment, subjects were first asked to explore the VR environment to get used to the interface and practice the simulated teleoperated task 10 to 15 times until they gained enough familiarity with the task. The goal of this process is to eliminate any learning effects on the continuous authentication performance.

Each subject was then asked to perform the task 5 times, while the following data were collected in real-time.

\begin{enumerate}
	\item Position of the center of the ring ($x,y,z$)
	\item Orientation of the ring ($Q_x,Q_y,Q_z,Q_w$, in quaternion)
	\item Force applied ($f_x,f_y,f_z$)
\end{enumerate}

All data were recorded at 60 Hz. In each trial, the app starts recording data when the user starts moving the ring and stops when the user pulls the ring out of the last peg.

Since the position and force data are highly correlated due to the influence of the virtual fixtures, we also collected velocity information from the gathered position data and use it to train the model. 

The state vector at time $t$ is then constructed as $s = (v_x,v_y,v_z,Q_x,Q_y,Q_z,Q_w,f_x,f_y,f_z)$.

\subsection{Gesture Segmentation}

In order to make the analogy between human motion and language, we first need to identify what corresponds to "words" in the teleoperation process, which is the operator's gesture. In this work, we segmented the entire process into 3 basic gestures as listed below.\\
1) Gesture 1: Transfer the ring toward next peg tip;\\
2) Gesture 2: Insert the ring;\\
3) Gesture 3: Pull out the ring.\\

The segmentation is based on the position of the ring and the corresponding status of the teleoperation process.

\subsection{Subject Demographics}
Our experimental results are based on our study involving 10 participants. The demographics of all participants are shown in Table \ref{tab:cont_exp_table1}. All experiments were conducted with the approval of the UW Institutional Review Board.

\begin{table}
	\begin{center}
		\caption{Subjects Demographics}
		\label{tab:cont_exp_table1}
		\scalebox{1}{
			\begin{tabular}{|c |c|}
				\hline
				Sample Size & 10  \\
				\hline
				Sex & 6 Females; 4 Males  \\
				\hline
				Age Range & 18 to 28 \\
				\hline
				%Age (Mean $\pm$ SD) & 26.5 $\pm$ 8.7\\
				%\hline
				Handedness & 2 Left; 8 Right\\
				\hline
				
			\end{tabular}
		}
	\end{center}

\end{table}    
	\section{HMM-based Continuous Operator Authentication}
In this paper, we use a Left-Right HMM\cite{rabiner1989tutorial} to model operators' gestures followed by a Token Passing algorithm \cite{young1989token} to concatenate the gesture models, thus achieving continuous authentication. The major advantage of using HMMs is that in the training phase, only positive samples are required. Also, HMMs are able to capture local dynamic properties of the operator's gestures\cite{roy2014hmm} which serves the purpose of continuous operator authentication. In the following sections, we will briefly review the Left-Right HMM Model and Token Passing Algorithm and demonstrate how we implement them for the proposed continuous authentication task.

\subsection{Left-Right HMM Model}
Left-Right HMMs have been widely used in speech recognition\cite{gales1998maximum, rabiner1993fundamentals, young1997htk}. They are used to model each word/phoneme separately. Compared to conventional HMMs, the Left-Right HMM offers non-emitting entry and exit states. They provide benefits by concatenating different words/phonemes together in real-time to achieve speech recognition. 

\begin{figure}
	\centering
	\includegraphics[width=0.6\columnwidth]{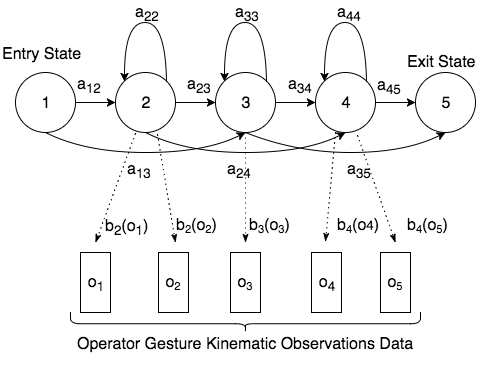}
	\caption{Left-Right Hidden Markov Model with Non-emitting State}
	\label{fig:chpt3_lrhmm}
\end{figure}

In this paper, we use left-right HMM to represent each gesture from an individual operator. The proposed Left-Right HMM structure is shown in Figure \ref{fig:chpt3_lrhmm}. Each state $i$ is associated with an emission probability distribution $b_i(o_t)$, which defines the probability of generating observation $o_t$ at time $t$. Additionally, the transition probability between each pair of states $i$ and $j$ is determined by transition probability $\{a_{ij}\}$. Furthermore, the entry (first) and exit (last) states of the proposed HMM are non-emitting. These two states are used to facilitate the concatenation between surgeme models as explained in more detail later. The other states are emitting states associated with emission probability distributions. The transition matrix is $N\times N$, where $N$ is the number of states. The sum of each row will be one except for the last row which is zero since no transition is allowed from the final state.

We assume that for each emitting state $i$ that the emission probability distribution is a Gaussian mixture as similar assumptions are frequently made with motion detection and speech recognition \cite{roy2014hmm, gales1998maximum, rabiner1993fundamentals, young1997htk}. For state $i$, the probability  $b_i(o_t)$ of generating observation $o_t$ is given by
\begin{equation}
	b_i(o_t)=\sum_{m=1}^{M_i}{c_{im}\mathcal{N}(o_t;\mu_{im},\Sigma_{im})}
\end{equation}

where $M_i$ is the number of Gaussian mixtures in state $i$, $c_{im}$ is the weight of the $m^\text{th}$ mixture and $\mathcal{N}(\cdot;\mu_{im},\Sigma_{im})$ is the probability density function of a multivariate Gaussian distribution with mean $\mu_{im}$ and covariance matrix $\Sigma_{im}$.

In the training phase, Baum-Welch re-estimation \cite{rabiner1989tutorial} is used. Segmented pieces of gesture sequences from each operator (as mentioned in section 5.2) are used as ground truth and the Baum-Welch algorithm is applied to obtain the maximum likelihood estimation of the model parameter state transition probability matrix $\{a_{ij}\}$ and emission probability distribution $b_i(o_t)$ for $i,j=1,...,N$.

\subsection{Token Passing Algorithm}
Given the observation sequence, to achieve gesture recognition and continuous authentication, the first step is to determine the hidden state sequence. This can be done using Viterbi Decoding Algorithm \cite{forney1973viterbi}. In this work, an alternative formulation of the Viterbi Algorithm called the Token Passing Algorithm\cite{young1989token} is used. It is able to realize single gesture recognition while simplifying concatenating gesture models for continuous operator authentication.

First, for the base case of single gesture recognition, the Token Passing algorithm works as follows. At each time frame $t$, the following algorithm is executed:\\

\textbf{for} t= 1 \textbf{to} T do\\
\tab\textbf{for each} state $i$ \textbf{do}\\
\tab\tab Pass a copy of the token in state $i$ to each connecting state $j$:\\
\tab\tab $\psi_{i \rightarrow j}(t) \leftarrow \psi_i(t-1)+log(a_{ij})+log(b_j(o(t)))$; \\
\tab \textbf{end}\\
\tab Discard the original tokens;\\
\tab\textbf{for each} state $j$ \textbf{do}\\
\tab\tab $\psi_j(t)=\max\limits_{i}\{\psi_{i \rightarrow j}(t)\}$ for each state $i$ connected to state $j$ \\
\tab \textbf{end}\\
\tab[0.5cm] \textbf{end}\\

where:

$\psi_i(t)$: the maximum log-likelihood of observing operation signal $o_{1:t}$ and being in state $i$ at time $t$;

$\psi_{i \rightarrow j}(t)$: the  log-likelihood of observing operation signal $o_{1:t}$ and a state transition from $i$ to $j$ at time $t$; 

$a_{ij}$: the state transition probability from state $i$ to state $j$;

$b_i(o_t)$: the emission probability of state $i$ given observation $o_t$.\\

In the Token Passing algorithm, it is assumed that each state $j$ of an HMM at time $t$ holds a single movable token that contains partial log-likelihood $\psi_j(t)$.

Let $o_{1:T}$ denote the gesture observation sequence with length $T$. $\psi_{max}(T|o_{1:T}, \textrm{ operator }i,\textrm{ gesture }j)$ denotes the log likelihood held by the remaining token at time $T$, given that the gesture model is from operator $i$ and gesture $j$. The gesture observation sequence $o_{1:T}$ can be recognized as operator $i$'s gesture $j$, given:
\begin{equation}
	(i, j)=arg\max\limits_{i,j}\{\psi_{max}(T|o_{1:T},\textrm{ operator }i,\textrm{ gesture }j)\}
\end{equation}

\begin{figure}[thpb]
	\centering
	\includegraphics[width=0.6\columnwidth]{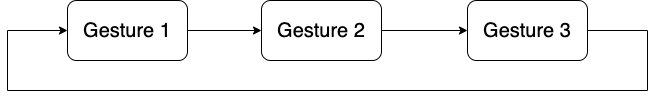}
	\caption{Gesture Grammar}
	\label{fig:chpt3_grammar}
\end{figure}

Next, in order to realize continuous operator authentication, individual gesture models need to be concatenated. Similar to human language, there is \emph{grammar} in the teleoperation task as well, which defines how the gestures can be connected. During the experiment, each subject first transfers the ring towards next peg tip (Gesture 1), inserts the ring until it reaches the back board (Gesture 2), pulls out the ring (Gesture 3) and then repeats these operations until the subject finishes the task. Therefore, in our proposed simulated teleoperated task, the grammar is defined as in Figure \ref{fig:chpt3_grammar}. 

\begin{figure}[thpb]
	\centering
	\includegraphics[width=0.7\columnwidth]{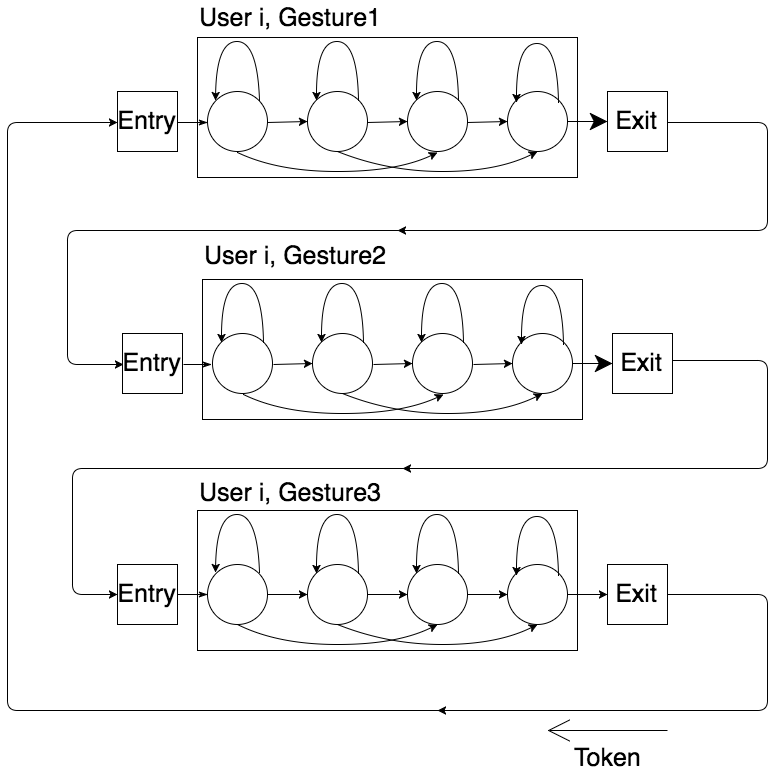}
	\caption{Token Passing Algorithm}
	\label{fig:chpt3_net}
\end{figure}

The structure of concatenating the gestures based on the grammar is shown in Figure \ref{fig:chpt3_net}. The non-emitting entry and exit states now work as \emph{glue}
to join gesture models together. At each time frame $t$, the following algorithm is executed for arbitrary observation sequence that potentially contains multiple gestures:

\textbf{for} t= 1 \textbf{to} T do\\
\tab\textbf{for each} state $i$ of gesture $k$ \textbf{do}\\
\tab\tab Pass a copy of the token in the state $i$ of the gesture $k$ to each connecting state $j$ of the gesture $k$ :\\
\tab\tab $\psi_{{(g_k,s_i)} \rightarrow {(g_k,s_j)}}(t) \leftarrow \psi_{(g_k,s_i)}(t-1)+log(a_{(k,ij)})+log(b_{(k,j)}(o(t)))$; \\
\tab\tab Pass a copy of the token in the state $i$ of the gesture $k$ to state $j$ of the gesture $l$ via non-emitting \\
\tab\tab exit and entry state based on the gesture grammar:\\
\tab\tab $\psi_{{(g_k,s_i)} \rightarrow {(g_l,s_j)}}(t) \leftarrow \psi_{(g_k,s_i)}(t-1)+log(a_{(k,i\rightarrow exit)})+log(a_{(l,entry\rightarrow j)})+log(b_{(l,j)}(o(t)))$; \\
\tab \textbf{end}\\
\tab Discard the original tokens;\\
\tab\textbf{for each} state $i$ of gesture $k$ \textbf{do}\\
\tab\tab $\psi_{(g_k,s_i)}=\max\limits_{j,l}\{\psi_{{(g_l,s_j)} \rightarrow {(g_k,s_i)}}(t)\}$ for each gesture $l$ connected to gesture $k$ based on the gesture grammar,\\
\tab\tab and each connected state $j$.\\
\tab \textbf{end}\\
\tab[0.5cm] \textbf{end}\\

where:

$\psi_{(g_k,s_i)}(t)$: the maximum log-likelihood of observing operation signal $o_{1:t}$ and being in the state $i$ of the gesture 

$k$ at time $t$.

$\psi_{{(g_k,s_i)} \rightarrow {(g_l,s_j)}}(t)$: the  log-likelihood of observing operation signal $o_{1:t}$ and a transition from the state $i$ of the 

gesture $k$ to the state $j$ of the gesture $l$ at time $t$; 

$a_{(k,ij)}$: the state transition probability for the gesture k from the state $i$ to the state $j$;

$a_{(k,i\rightarrow exit)}$: the state transition probability for the gesture k from the state $i$ to the non-emitting exit state;

$a_{(k,entry\rightarrow j)}$: the state transition probability for the gesture k from the non-emitting entry state to the state $j$;

$b_{(k,i)}(o_t)$: the emission probability of the gesture $k$, state $i$ given observation $o_t$.\\

For connected gesture recognition, besides the overall log-likelihood, we also want to know the best matching gesture sequence. Tokens are assumed to hold a path identifier as well as the path log-likelihood. The path identifier is used to record gesture boundary information which will be called Gesture Link Record (GLR). At each time $t$, extra steps shown as follows are taken in addition to the individual gesture recognition algorithm listed above:\\

\textbf{for each} token entered EXIT state at time t do\\
\tab[0.6cm] create a new GLR containing:\\
\tab[1.2cm] $<$token contents, $t$, identity of emitting gesture$>$;\\
\tab[0.6cm] change the path identifier of the token to point to this \\
\tab[0.6cm] new record\\
\tab[0.3cm] \textbf{end}\\

By doing so, potential gesture boundaries are recorded in a linked list, and on completion at time $T$, the path identifier held in the token with the largest log-likelihood can be used to trace back through the linked list to find the best gesture sequence and the corresponding gesture boundary locations.

\subsection{Continuous Operator Authentication}

To accomplish continuous operator authentication, we put an additional constraint on the aforementioned gesture recognition scheme by mandating that consecutive gestures must come from the same operator.

At time $t$, given the observation from the sample window with width $T$ as $O_{t-T:t}$, operator recognition is done by solving $\psi_{max}(t)$ and checking the gesture labeling $l_{t-T:t}$. The observation sequence will be recognized as operated by user $i$ if
\begin{equation}
	l_{t_0}\in L_i, t_0=t-T,...,t
\end{equation}

where $i$ is the operator ID and $L_i$ is the corresponding gesture label set for the $i^{th}$ user.

	\section{Results}
\subsection{Authentication Result}
In this work, we used the leave-one-trial-out cross-validation strategy to train and test the proposed method. 

First, in the training phase, the gesture models from each subject are trained based on the segmented individual gesture pieces in those training trials. In this way, 3N subject gesture models were obtained where 3 is the number of gestures and N is the number of the subjects.

We varied the hyperparameters of the HMM to test the corresponding continuous authentication performance. We varied the number of states from 3 to 6 and the number of mixtures in each state from 1 to 3. We also tested moving sample windows with widths of 5 seconds, 3 seconds, and 1 second. 

Let $L_{window}$ denote the number of observations contained by a moving sample window. As the kinematic data for the VR experiment is recorded at 60 Hz, the tested 5s, 3s, 1s moving sample windows contain 300, 180, and 60 observations respectively. $L_{i,j}$ denote the total number of observations contained by the $j^{th}$ trial from subject $i$. The continuous authentication starts when there is at least $L_{window}$ observations. Each time the moving sample window is shifted by 1 observation, hence the size of the overlap between two consecutive sample windows is $L_{window}-1$. As a consequence, there are $L_{i,j}-L_{window}+1$ sample windows evaluated for the testing trial. The HMM models generate a predicted subject label for each sample window. Therefore, the performance analysis is equivalent to a multi-class classification on the sample windows from all the trials. The accuracy is calculated as:

\begin{equation}
Accuracy = \frac{N_{hit}}{N_{tot}}
\end{equation}

where:

$N_{hit}$: the total number of sample windows where the subject is correctly classified in all trials.

$N_{tot}$: the total number of sample windows in all trials.

\begin{figure}
	\begin{subfigure}{0.3\columnwidth}
		\centering
		\includegraphics[width=0.99\columnwidth]{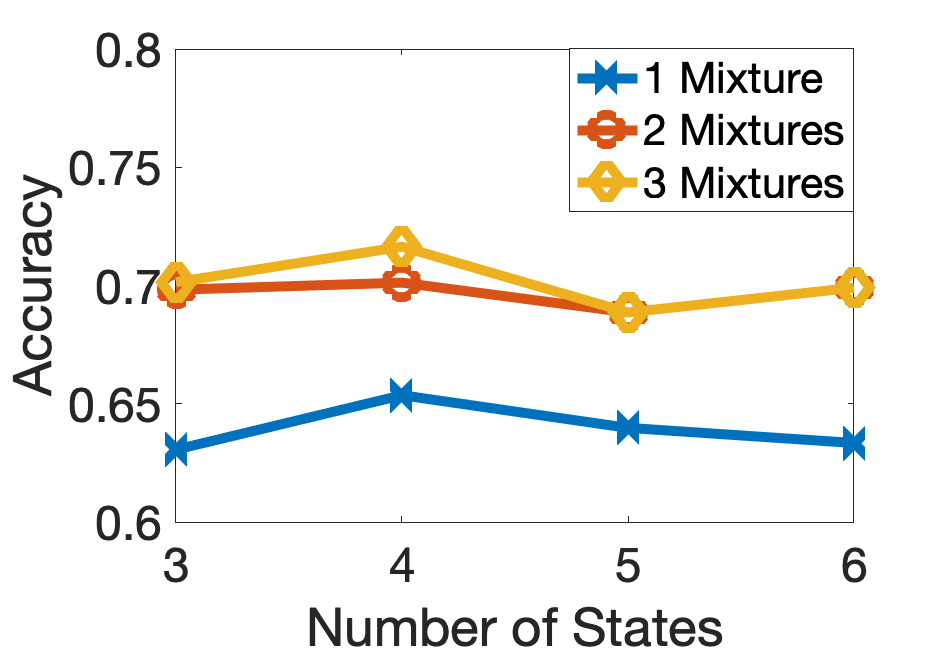}
		\caption{1-Second Sample Window}
	\end{subfigure}
	\begin{subfigure}{0.3\columnwidth}
		\centering
		\includegraphics[width=0.99\columnwidth]{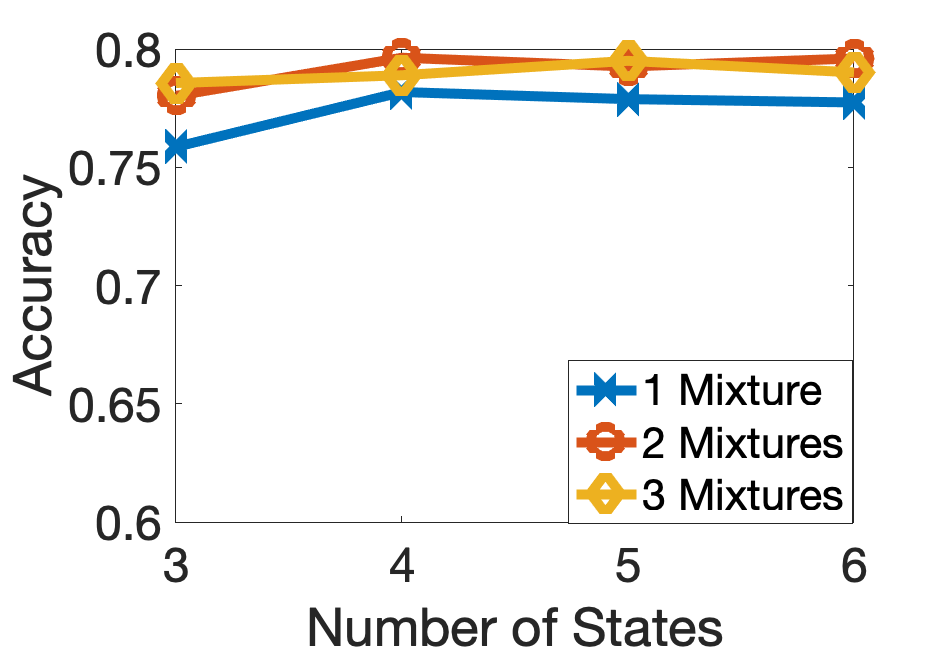}
		\caption{3-Second Sample Window}
	\end{subfigure}
	\begin{subfigure}{0.3\columnwidth}
		\centering
		\includegraphics[width=0.99\columnwidth]{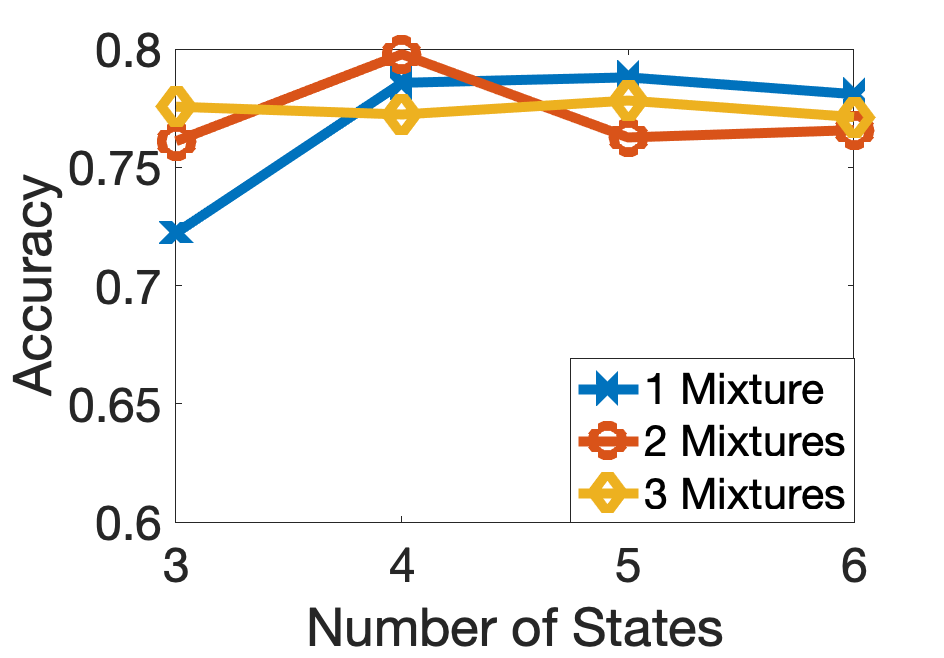}
		\caption{5-Second Sample Window}
	\end{subfigure}
	\caption{Continuous Authentication Accuracy with 1-Second(a), 3-Second(b) and 5-Second(c) Sample Window}
	\label{fig:chpt3_1s}
\end{figure}

The corresponding results with different sample window width and hyperparameters are shown in Figure \ref{fig:chpt3_1s}.

We found that when we choose the hyperparameter of the HMM which models each gesture as 4 states with 2 Gaussian mixtures for each state, we were able to achieve the best authentication accuracy. We set the hyperparameter as 4 states with 2 mixtures in the following context. The detailed confusion matrices for the VR experiment with 5s, 3s, and 1s sample window are shown in Table \ref{tab:app_vr_conf_5s}-\ref{tab:app_vr_conf_1s} in the Appendix respectively. The authentication accuracy, macro average precision, and macro average recall are listed in Table \ref{tab:chpt3_acc}.

\begin{table}[h]
	\begin{center}
		\caption{Continous Authentication Performance with Multiple Sample Window Width for the VR Experiments}
		\label{tab:chpt3_acc}
		\scalebox{1}{
			\begin{tabular}{|c |c| c |c|}
				\hline
				Window Width & 5 sec & 3 sec & 1 sec  \\
				\hline
				Accuracy & 79.78\% & 79.63\% & 70.12\%  \\
				\hline
				Avg. Precision & 79.48\% & 79.75\% & 70.71\%  \\
				\hline
				Avg. Recall & 82.54\% & 82.12\% & 73.12\%  \\
				\hline                
			\end{tabular}
		}
	\end{center}    
\end{table}    

When we used a sample window width of 5 seconds, we were able to authenticate the operator in realtime with almost 80\% accuracy, average precision, and recall. With a 1-second sample window, we achieved above 70\% continuous authentication accuracy, average precision, and recall. It shows that this method works and is promising for continuous authentication performance. 

\subsection{Simulated Impersonation Attack Resistance}
We simulated an impersonation attack in the following steps. First, we picked two subjects (represented as User 1 and User 2 in the following context) and used the leave-one-trial-out strategy to train the model for the gestures of each subject. In the testing phase, instead of using the remaining testing trial to examine the authentication performance, we split the test trial from User 1 and User 2 into 2 half pieces and concatenate User 1's first piece to User 2's second piece and vice versa. In this way, we generated two artificial teleoperation process observation sequences, where User 2 impersonates User 1 during the second half of the teleoperation task and vice versa. Compared to actual malicious impersonation attacks, we anticipate that the simulated impersonation attack is more difficult to detect. As in the simulated impersonation attack, both users were performing benign operations, hence the operations were similar. We anticipate the differences between malicious operations and benign operations are more significant than the differences among benign operations.

\begin{figure}[thpb]
	\centering
	\includegraphics[width=1\columnwidth]{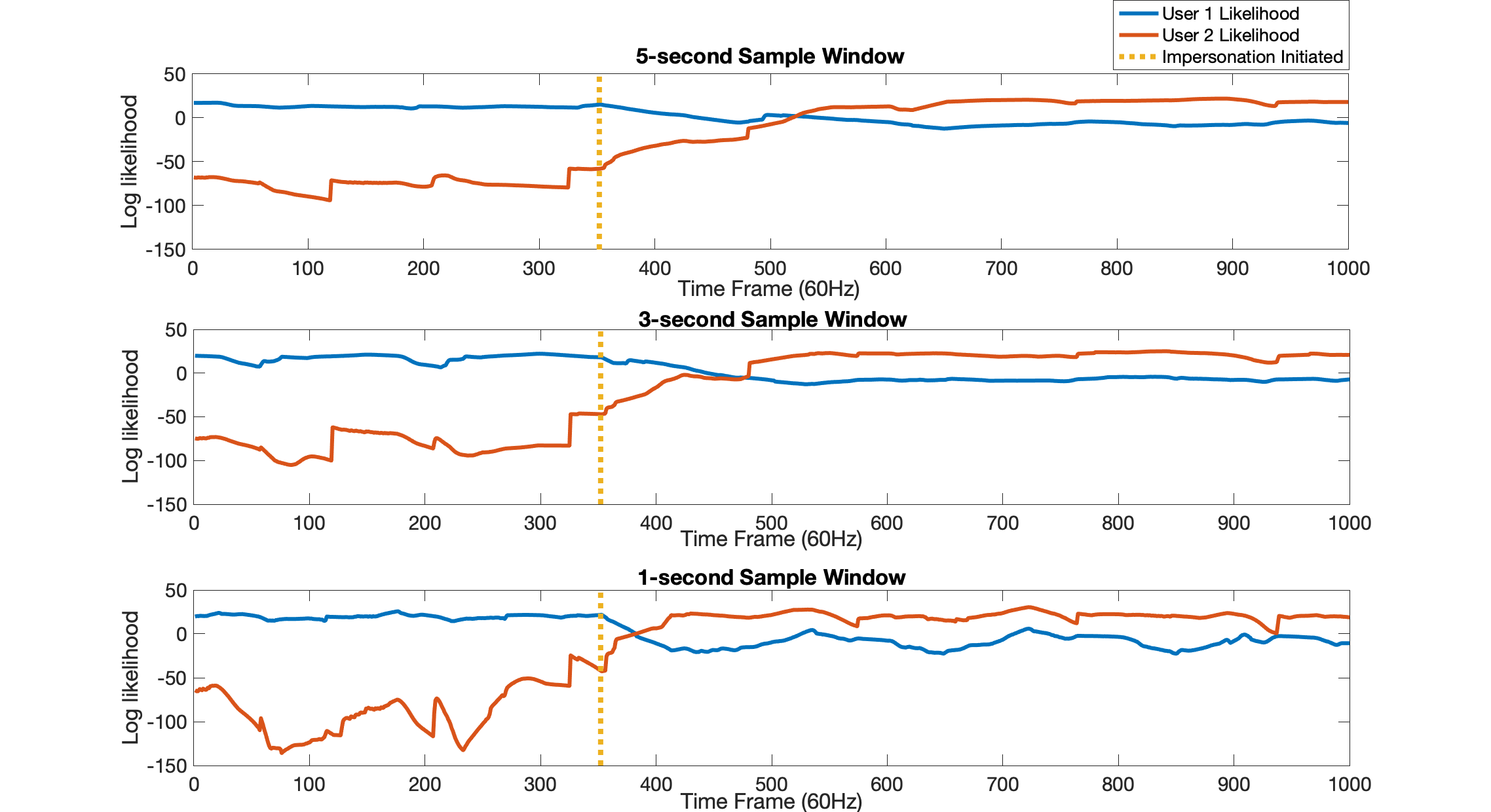}
	\caption{Simulated Impersonation Attack For the VR Experiment}
	\label{fig:chpt3_attack}
\end{figure}

Figure \ref{fig:chpt3_attack} shows one of the results in detail. The blue and red lines represent the likelihood that the data within the corresponding sample window is operated by User 1 or User 2 respectively. The orange dashed line is the point where the simulated impersonation attack takes place. First, this method is able to detect the impersonation attack as the likelihood of the original user drops significantly after the simulated attack launches. Moreover, we notice that with a longer sample window, it is easier to distinguish two operators as the differences between the likelihood of the two users are significantly larger at various points when the size of the sample window increases. On the other hand, the response time for the continuous authentication system to detect the impersonation attack increases when the sample window size becomes wider. Denoting the response time as the time between the time of the impersonation attack and the likelihood cross point between two users, the average response time is shown in Table \ref{tab:chpt3_response}.

\begin{table}[h]
	\begin{center}
		\caption{Average Response Time to Impersonation Attack with Multiple Sample Window Widths}
		\label{tab:chpt3_response}
		\scalebox{1}{
			\begin{tabular}{|c |c| c |c|}
				\hline
				Window Width & 5 sec & 3 sec & 1 sec  \\
				\hline
				
				Response Time & 2.35 sec & 1.51 sec & 0.49 sec  \\
				\hline                
			\end{tabular}
		}
	\end{center}    
\end{table}    

Therefore, when choosing the size of the sample window, there is a tradeoff between the accuracy of the continuous authentication and response time to attacks. A wider sample window is able to generate more stable continuous authentication accuracy, however, it takes more time to respond to the attack. On the other hand, shorter sample windows offer the ability to react quickly to impersonation attacks, but the authentication accuracy is less stable.

\subsection{Authentication Performance on Surgical Robotic System}
We tested the continuous operator authentication method on JHU-ISI Gesture and Skill Assessment Working Set (JIGSAWS)\cite{gaojhu} and explored the authentication performance of the developed method on a surgical telerobotic system. In JIGSAWS, the dataset is obtained through the da Vinci Surgical Robot, where subjects were asked to perform several surgery tasks. Kinematic data of three basic surgical tasks (suturing, needle passing, and knot tying) performed by 8 study subjects are included. The kinematic data were recorded at 30 Hz. Each task was performed 3-5 trials for each subject. Each teleoperation process is manually labeled as a sequence of surgical gestures. The definition of each gesture\cite{gaojhu} is presented in Table \ref{tab:gesture}.

\begin{table}[h]
	\caption{Surgical Gesture Definition\cite{gaojhu}}
	\label{tab:gesture}
	\begin{center}
		\scalebox{1}{
			\begin{tabular}{|c|l|}
				\hline
				Gesture Index & Surgeme Definition\\
				\hline
				G1 & Reaching for needle with right hand \\
				\hline
				G2 & Positioning needle \\
				\hline
				G3 & Pushing needle through tissue \\
				\hline
				G4 & Transferring needle from left to right \\
				\hline
				G5 & Moving to center with needle in grip \\
				\hline
				G6 & Pulling suture with left hand \\
				\hline
				G7 & Pulling suture with right hand \\
				\hline
				G8 & Orienting needle \\
				\hline
				G9 & Using right hand to help tighten suture \\
				\hline
				G10 & Loosening more suture \\
				\hline
				G11 & Dropping suture at end and moving to end points \\
				\hline
				G12 & Reaching for needle with left hand \\
				\hline
				G13 & Making C loop around right hand \\
				\hline
				G14 & Reaching for suture with right hand \\
				\hline
				G15 & Pulling suture with both hands \\
				\hline

			\end{tabular}
		}
	\end{center}
\end{table}

Given the observation kinematic data and gesture labeling, we segmented the teleoperated surgical process into gesture pieces and use that as ground truth to train the corresponding subject's gesture HMM models. To unify the data properties for each subject, we focused on those subjects with complete 5 trial kinematic and labeling data for each surgical task. By doing so, we obtained 7 valid subjects for the suturing task (indexed as "B", "C", "D", "E", "F", "G", "I" in JIGSAWS), 2 valid subjects for the needle passing task (indexed as "C", "D" in JIGSAWS) and 5 valid subjects for the knot tying task (indexed as "C", "D", "E", "F", "G" in JIGSAWS).

\begin{figure}[thpb]
	\centering
	\includegraphics[width=0.8\columnwidth]{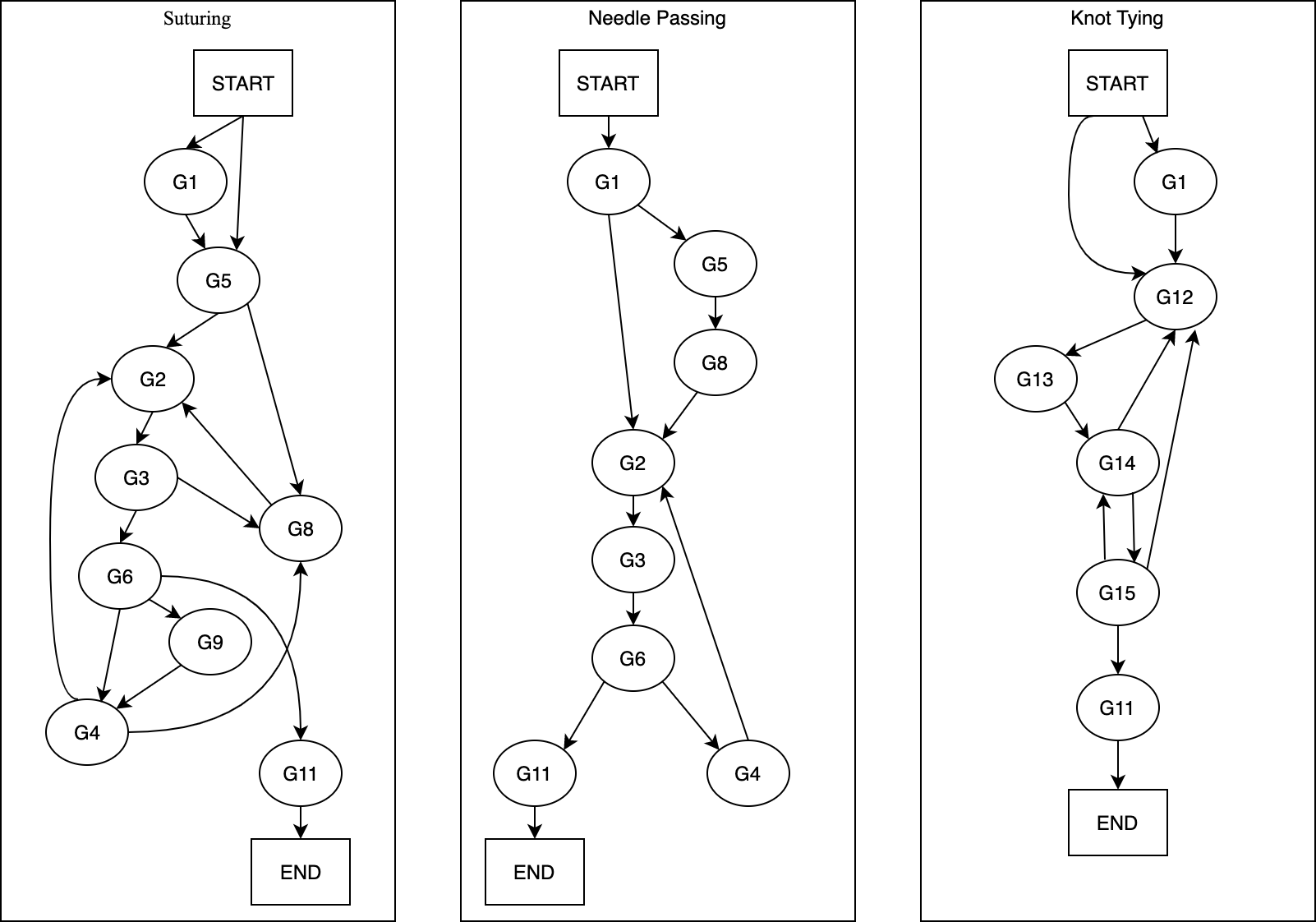}
	\caption{Grammar graph for suturing (left), needle passing (center), and knot tying (right)\cite{ahmidi2017dataset} }
	\label{fig:surg_grammar}
\end{figure}

For each surgical task, we kept the leave-one-trial-out setting to train and test the continuous operator authentication. In the training phase, we obtained $MN$ gesture models, where $M$ is the number of gesture types and $N$ is the number of subjects we tested. Table \ref{tab:chapter6_data} shows the detailed subject and gesture indices for all 3 surgical tasks. We tested the sample window with a width of 5 seconds, 3 seconds, and 1 second. We use the same setting as discussed in the previous section to evaluate the continuous authentication accuracy. The grammar of the surgical gesture\cite{ahmidi2017dataset} for each task is presented in Figure \ref{fig:surg_grammar}, which defines how the gesture models are connected for each surgical task.

\begin{table}[h]
	\begin{center}
		\caption{Subjects Evaluated and Gestures In the Corresponding JIGSAWS Tasks}
		\label{tab:chapter6_data}
		\scalebox{1}{
			\begin{tabular}{| c |c |c| c |c|}
				\hline
				& Number of Subjects & Subject Indices & Number of Gestures & Gesture Indices  \\
				\hline
				Suturing & 7 & B,C,D,E,F,G,I & 9 & G1,G2,G3,G4,G5,G6,G8,G9,G11  \\
				\hline                
				Needle Passing & 2 & C,D & 8 & G1,G2,G3,G4,G5,G6,G8,G11  \\
				\hline                
				Suturing & 5 & C,D,E,F,G & 6 & G1,G11,G12,G13,G14,G15  \\
				\hline                
			\end{tabular}
		}
	\end{center}    
\end{table}

We then obtain the following continuous authentication result as shown in Table \ref{tab:chpt3_acc_jigsaw}. The detailed confusion matrices for all sample window widths and tasks are listed in Appendix A.2.

\begin{table}[thpb]
	\begin{center}
		\caption{Continous Authentication Performance with Multiple Sample Window Width for the JIGSAWS dataset}
		\label{tab:chpt3_acc_jigsaw}
		\includegraphics[scale=0.45]{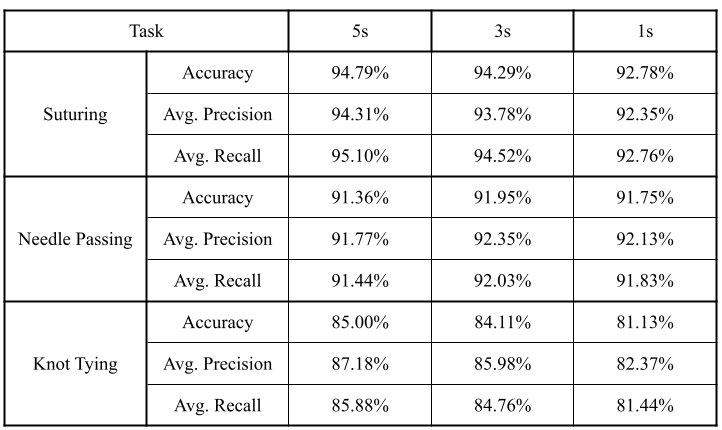}
		
	\end{center}    
\end{table}    

From these results, we found that even with 1-second
observation sequence, the continuous operator authentication accuracy to detect the subject is above 80\% for all three tasks. This shows that the developed continuous authentication method also works for teleoperated robotic surgeries.

We also conducted a simulated impersonation attack using the same setup as discussed in section 6.2. We performed the analysis on Subjects C and D as they are included in all 3 surgical tasks.

\begin{figure}[thpb]
	\centering
	\includegraphics[width=1\columnwidth]{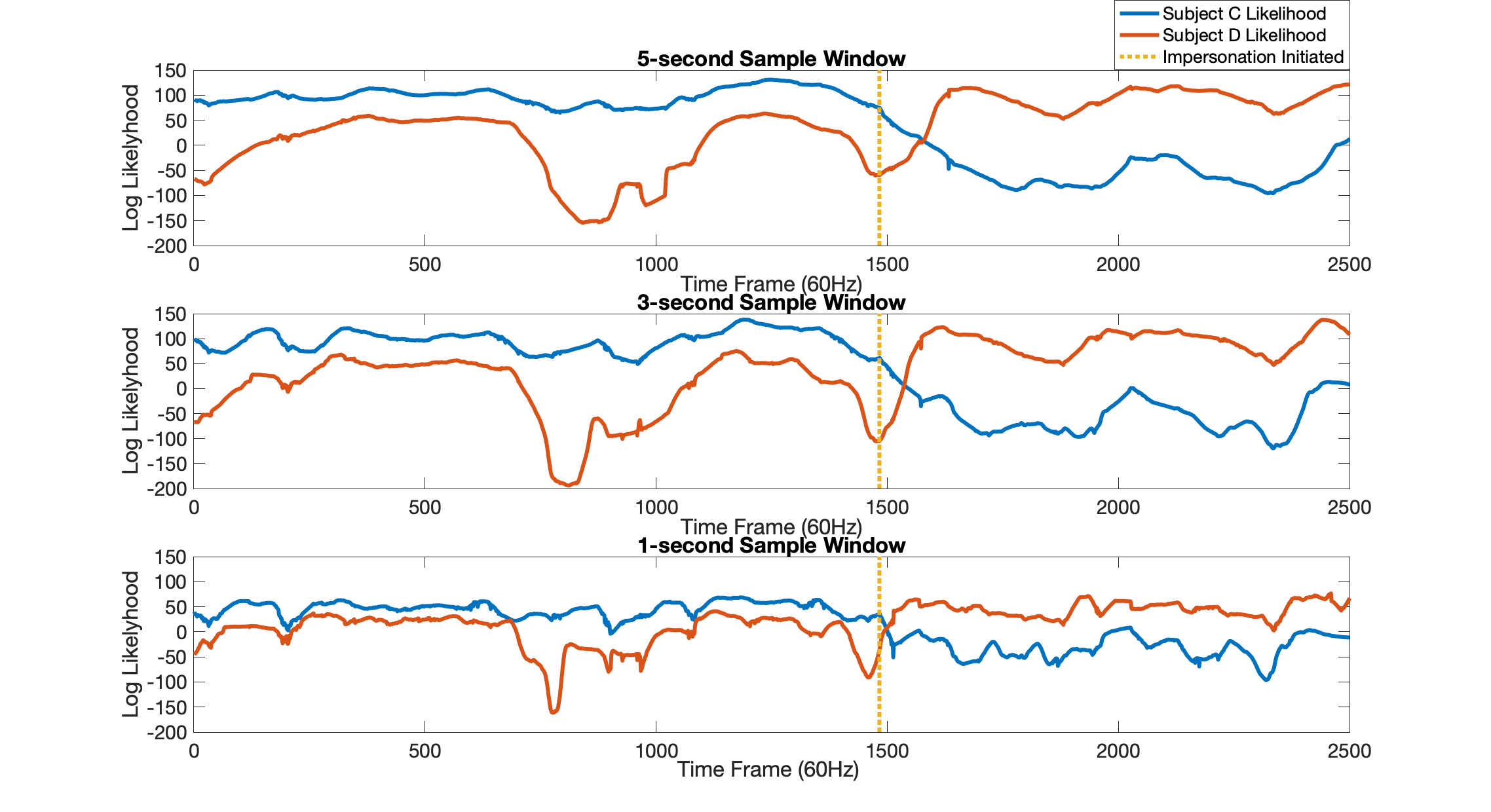}
	\caption{Simulated Impersonation Attack on the Suturing Task}
	\label{fig:chpt6_imp_st}
\end{figure}

\begin{figure}[thpb]
	\centering
	\includegraphics[width=1\columnwidth]{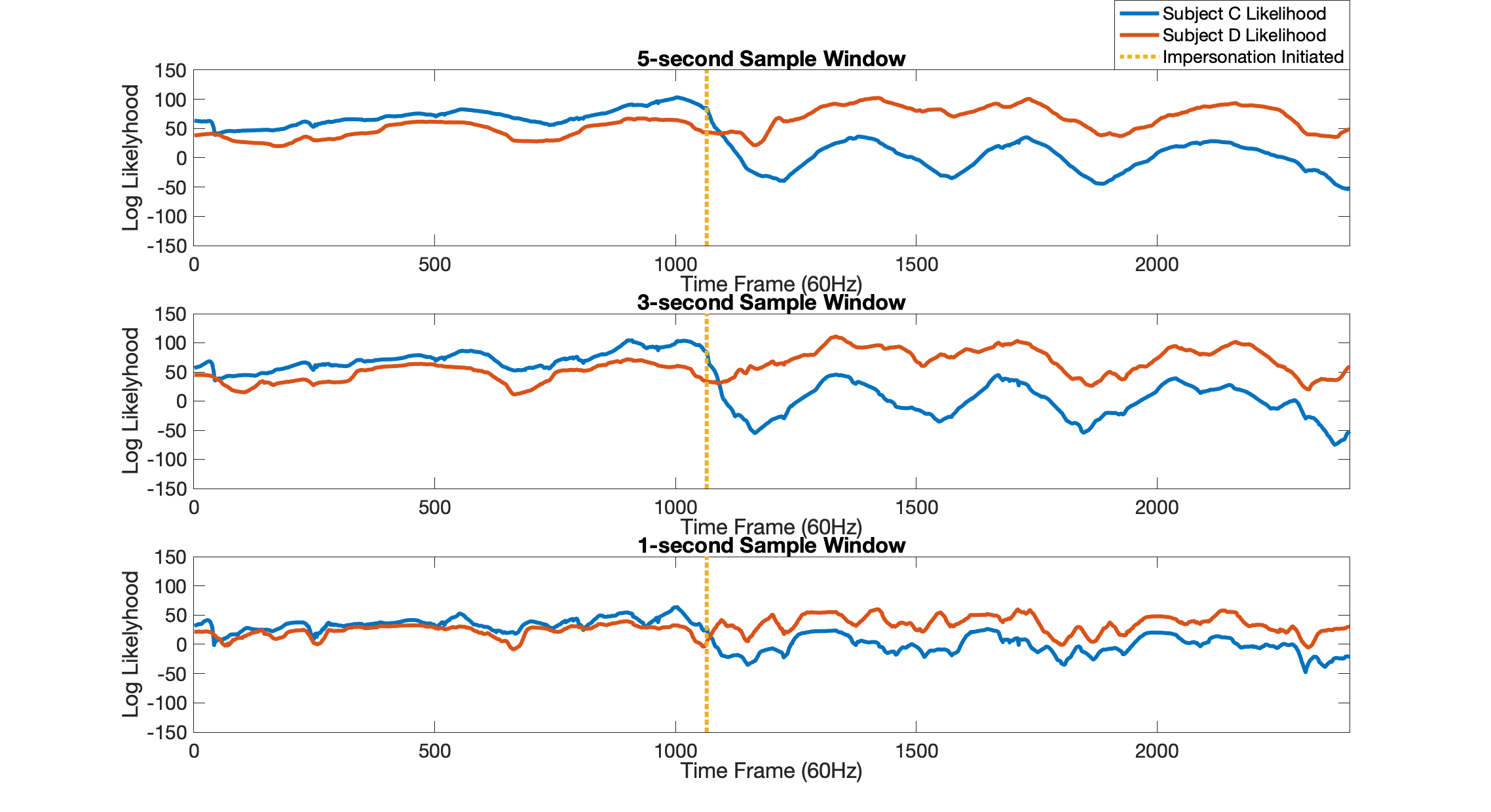}
	\caption{Simulated Impersonation Attack on the Needle Passing Task}
	\label{fig:chpt6_imp_np}
\end{figure}

\begin{figure}[thpb]
	\centering
	\includegraphics[width=1\columnwidth]{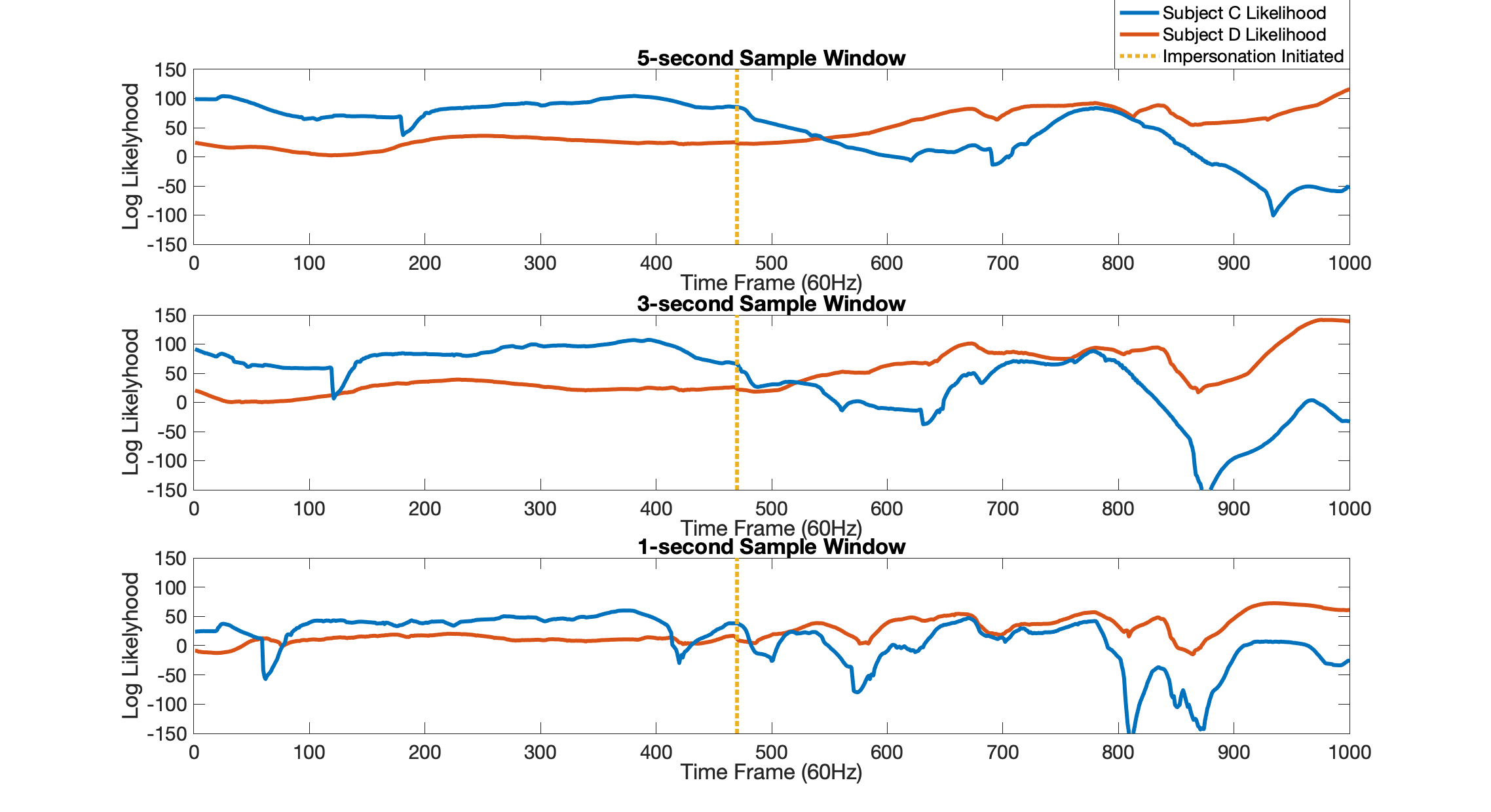}
	\caption{Simulated Impersonation Attack on the Knot Tying Task}
	\label{fig:chpt6_imp_kt}
\end{figure}

Figure \ref{fig:chpt6_imp_st}-\ref{fig:chpt6_imp_kt} show one of the results of each surgical task in detail. The average response time for each surgical task is shown in Table \ref{tab:chpt6_jigsaws_response}. The results confirm that for all 3 surgical tasks, the proposed method is able to detect the simulated impersonation attack. Additionally, similar to the VR experiment, the tradeoff between accuracy and response time is demonstrated as well. A larger sample window can generate more stable authentication accuracy with a longer response time to the attack, while a smaller sample window can generate less stable authentication accuracy with a faster response to the attack.


\begin{table}[h]
	\begin{center}
		\caption{Average Response Time to Impersonation Attack for All JIGSAWS Surgical Tasks}
		\label{tab:chpt6_jigsaws_response}
		\scalebox{1}{
			\begin{tabular}{|c |c| c |c|}
				\hline
				Window Width & 5 sec & 3 sec & 1 sec  \\
				\hline
				
				Suturing & 2.99 sec & 1.73 sec & 0.53 sec  \\
				\hline                
				Needle Passing & 2.55 sec & 1.56 sec & 0.62 sec  \\
				\hline                
				Knot Tying & 2.35 sec & 1.26 sec & 0.42 sec  \\
				\hline                
			\end{tabular}
		}
	\end{center}    
\end{table}  
	\section{Discussion}

In this section, we will discuss several limitations of this paper and also raise possible extensions of our approach.

\textbf{Inexperienced Experimental Subjects} In the VR simulated teleoperation experiment, most subjects had never interacted with VR and/or haptic input devices. Although we conducted a training session to familiarize them with the system, we still noticed that there was a learning effect during the data collection, whereby the subject became better in handling and operating the system. Also, in some cases, subjects were less patient towards the end of the experiment, which influenced their motion to deviate from the original model. All these factors might have undermined the authentication performance result in our experiment. However, in most real-life applications, genuine operators are usually well trained and have sufficient familiarity and experience with the teleoperated system. Therefore, it is more likely that the operator has a unique operating `pattern' which makes it easier to accomplish the continuous operator authentication.

On the other hand, when dealing with the JIGSAWS dataset, we were able to achieve better continuous operator authentication accuracy compared to the VR simulated experiment as some of the subjects have previously trained to operate the surgical robot. Moreover, in the original JIGSAWS dataset, all the subjects are categorized as 3 skill levels based on their robotic surgical experience (Expert: more than 100 hours, Intermediate: between 10 and 100 hours, Novice: less than 10 hours). Table \ref{tab:expert} represents the average continuous authentication accuracy among experts and non-experts for all 3 surgical tasks. The result shows that our approach can achieve better continuous authentication performance when dealing with expert subjects. It confirms our assumption that it would be easier to extract more unique features from the operator and achieve better performance when the operator has sufficient experience with the teleoperated system. 

\begin{table}[h]
	\begin{center}
		\caption{Continous Authentication Performance with Different Skill Level}
		\label{tab:expert}
		\includegraphics[scale=0.45]{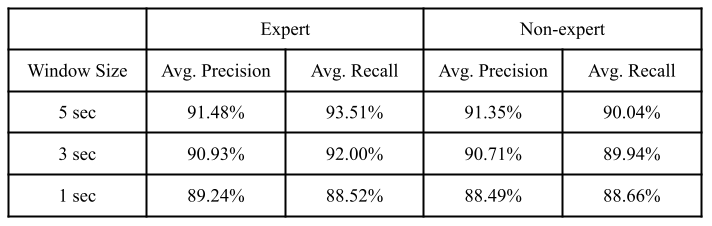}
	\end{center}    
\end{table}    

\textbf{Teleoperation Tasks Involve Intended Operator Switches} In some teleoperation tasks, such as robotic-assisted surgeries, there are intentional operator switches. A single operator model is not sufficient to continuously authenticate all the collaborating operators. In this case, instead of using a single operator model, we can develop an allowlist of operator models to capture all the authenticated operators. During the continuous authentication process, the system can authenticate the current operator as long as data within the current sample window gets authenticated by one of the operator models in the allow list.

\textbf{Teleoperation Task Complexity} In this work, our experimental results were based on relatively simple (VR simulation) and well-structured (basic surgical tasks) teleoperation tasks. In real-life scenarios, some teleoperation tasks are more complex and less structured. It would be interesting to explore whether our approach is able to achieve good continuous authentication performance in these applications. We anticipate that better gesture segmentation, as well as gesture grammar, is needed to better generalize these types of teleoperation processes. 

\textbf{Authentication Performance Under Various Conditions} In the proposed experimental study, we were only able to use data collected within a single-day for each subject and the subjects all exhibit normal cognitive condition. In\cite{gallagher2011persistent}\cite{white2013quantitative}, how teleoperators' behavior patterns vary over time and under different cognitive condition were investigated. Therefore, teleoperators' gestures vary over the long-term and different cognitive conditions may affect the continuous authentication performance. 

For example, within each teleoperation procedure, the operator's fatigue and pressure under time constrain may change an operator's behavior pattern. Additionally, the operators' experise level will improve over time and thus change how they interact with the remote robot. It is worth further investigation on how these factors will affect the continuous operator authentication accuracy.

A possible way to address this issue is to develop a mechanism to adaptively update the operator's authentication model based on the new test data. Every time a verified genuine operation process is conducted, gesture sequences can be obtained either manually or by using automatic surgical gesture segmentation as proposed in \cite{lin2006towards, tao2013surgical, lea2016temporal, lea2016segmental,lea2017temporal, ahmidi2017dataset}. The model can then be updated by using the new set of gestures. This will help the authentication model to capture the operator's behavior pattern under different operation conditions, such as fatigue or high-stress level. It also allows the model to accommodate for the operator's long term variation due to evolving skill level or aging.  

\textbf{Replay Attacks} One possible way for an attacker to bypass the proposed continuous authentication is through replay attacks. If an attacker can record a benign kinematic operation command sequence from an authorized operator, it is possible for the attacker to edit sequences to smoothly transition into and between replay attacks. Such types of attacks can potentially cause damage as well. However, as the operation command sequence is generated by an authorized genuine operator, and the transition between each replay has been smoothed out, the proposed continuous authentication will not be able to detect them. Nevertheless, the human-in-the-loop nature of teleoperation processes allows us to mitigate this problem. Since humans are not perfect machines, it is almost impossible for any human to generate exactly the same kinematic operation command via a control console. A command sequence history can be monitored to raise an alert when an exact or near-duplicate command sequence (i.e. attackers may add small noise to each replay) is observed. This will allow us to overcome the problem of replay attacks against the proposed continuous authentication method.
	\section{Conclusion}
In this paper, we develop a continuous and real-time operator authentication method by making an analogy between human motion and human language (gesture to word and operation process to sentence). We use HMMs to model each operator's gestures and then concatenate them by using the Token Passing Algorithm based on a predefined operation grammar to achieve continuous authentication. We built a VR simulated environment and conducted a human subject experiment where the subjects conducted a simulated teleoperation task within the VR environment. We also tested our approach on a teleoperated surgical process as we used the JIGSAWS dataset and explored its continuous authentication performance. Our experimental results indicate that the developed continuous teleoperator authentication method works and is able to achieve above 70\% accuracy rate for the VR simulated teleoperation task and 81\% accuracy rate for JIGSAWS surgical tasks with as short as a 1-second sample window. Moreover, we further examined the continuous authentication system resistance to impersonation attack and demonstrated that our approach is able to detect impersonation attacks with short response time.

	\begin{acks}
		This material is based upon work supported by the National Science Foundation under Grant No. CNS-1329751. Any opinions, findings, and conclusions or  recommendations expressed in this material are those of the author(s) and do not necessarily reflect the views of the National Science Foundation.
	\end{acks}
	
	\bibliographystyle{ACM-Reference-Format}
	\bibliography{uwthesis}
	
	\appendix
\section{Confusion Matrices}
\subsection{VR Experiements}
\begin{table}[thpb]
	\begin{center}
		\caption{Confusion Matric for the VR Experiment with 5-second Sample Window}
		\label{tab:app_vr_conf_5s}
		\includegraphics[scale=0.33]{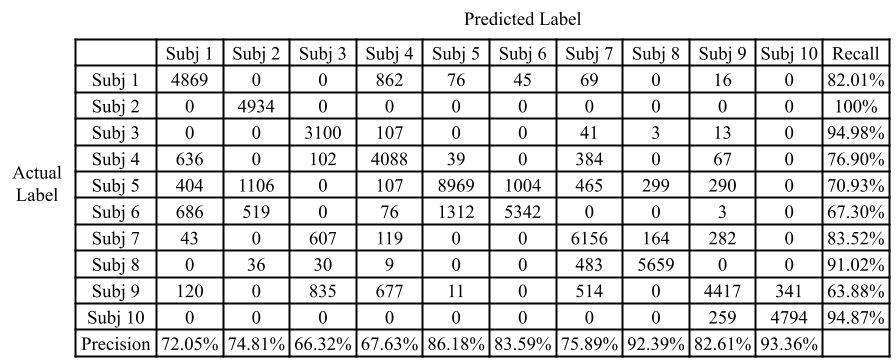}
		
	\end{center}    
\end{table}  
\begin{table}[thpb]
	\begin{center}
		\caption{Confusion Matric for the VR Experiment with 3-second Sample Window}
		\label{tab:app_vr_conf_3s}
		\includegraphics[scale=0.33]{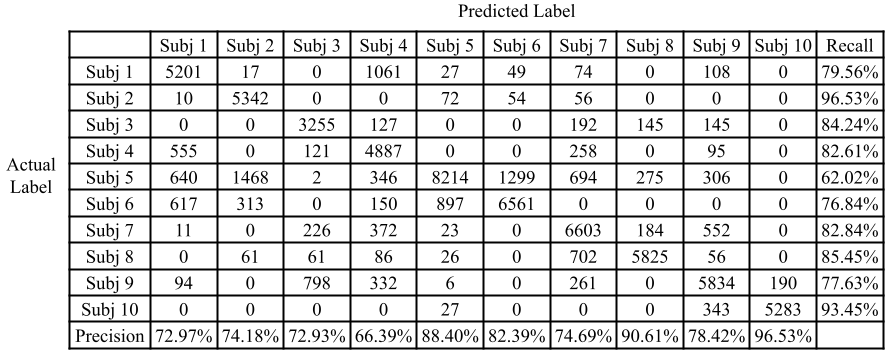}
		
	\end{center}    
\end{table}  
\begin{table}[thpb]
	\begin{center}
		\caption{Confusion Matric for the VR Experiment with 1-second Sample Window}
		\label{tab:app_vr_conf_1s}
		\includegraphics[scale=0.33]{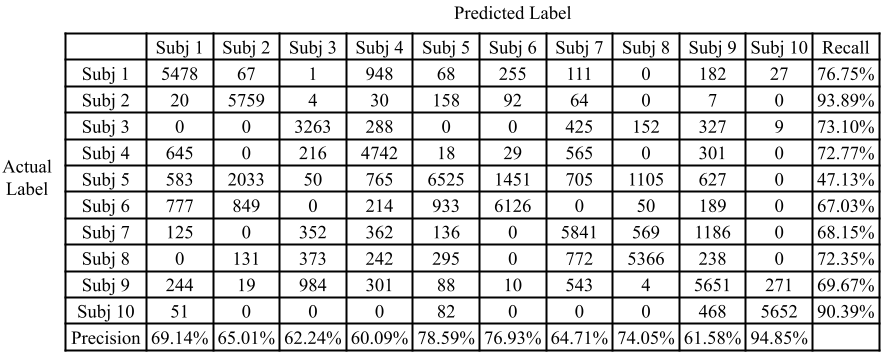}
		
	\end{center}    
\end{table}  

\subsection{JIGSAWS Dataset}
\begin{table}[thpb]
	\begin{center}
		\caption{Confusion Matric for the JIGSAWS Suturing Task with 5-second Sample Window}
		\label{tab:app_st_conf_5s}
		\includegraphics[scale=0.33]{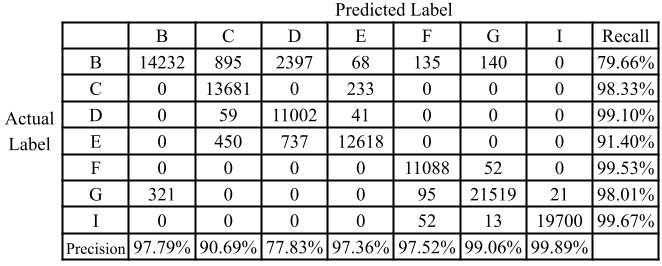}
		
	\end{center}    
\end{table}  
\begin{table}[thpb]
	\begin{center}
		\caption{Confusion Matric for the JIGSAWS Suturing Task with 3-second Sample Window}
		\label{tab:app_st_conf_3s}
		\includegraphics[scale=0.33]{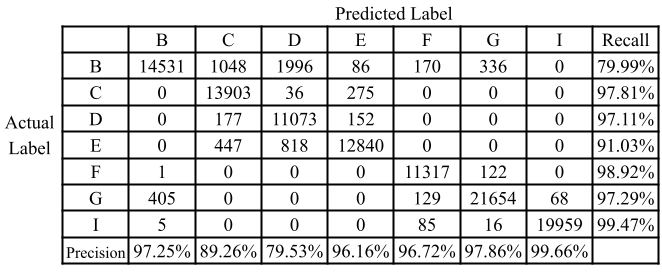}
		
	\end{center}    
\end{table}  
\begin{table}[thpb]
	\begin{center}
		\caption{Confusion Matric for the JIGSAWS Suturing Task with 1-second Sample Window}
		\label{tab:app_st_conf_1s}
		\includegraphics[scale=0.33]{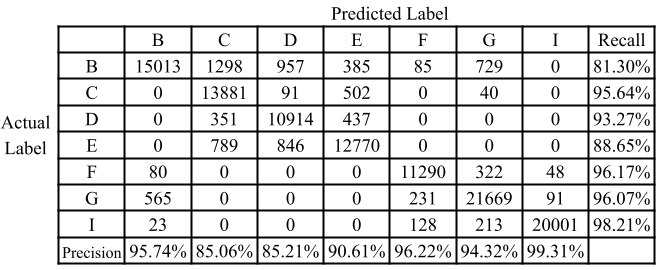}
		
	\end{center}    
\end{table}
\begin{table}[thpb]
\begin{center}
	\caption{Confusion Matric for the JIGSAWS Needle Passing Task with 5-second Sample Window}
	\label{tab:app_np_conf_5s}
	\includegraphics[scale=0.33]{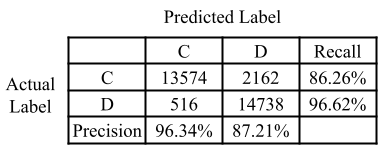}
	
\end{center}    
\end{table}  
\begin{table}[thpb]
\begin{center}
	\caption{Confusion Matric for the JIGSAWS Needle Passing Task with 3-second Sample Window}
	\label{tab:app_np_conf_3s}
	\includegraphics[scale=0.33]{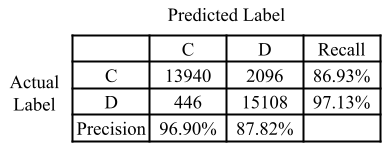}
	
\end{center}    
\end{table}  
\begin{table}[thpb]
\begin{center}
	\caption{Confusion Matric for the JIGSAWS Needle Passing Task with 1-second Sample Window}
	\label{tab:app_np_conf_1s}
	\includegraphics[scale=0.33]{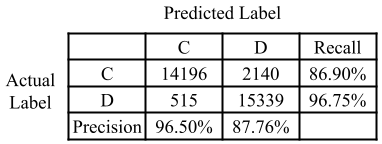}
	
\end{center}    
\end{table}
\begin{table}[thpb]
	\begin{center}
		\caption{Confusion Matric for the JIGSAWS Knot Tying Task with 5-second Sample Window}
		\label{tab:app_kt_conf_5s}
		\includegraphics[scale=0.33]{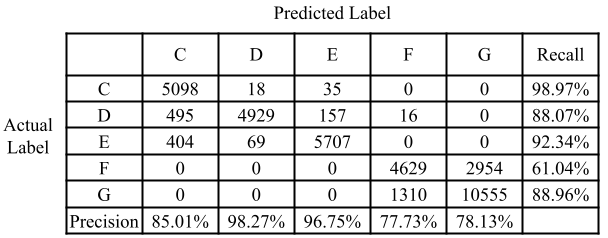}
		
	\end{center}    
\end{table}  
\begin{table}[thpb]
	\begin{center}
		\caption{Confusion Matric for the JIGSAWS Knot Tying Task with 3-second Sample Window}
		\label{tab:app_kt_conf_3s}
		\includegraphics[scale=0.33]{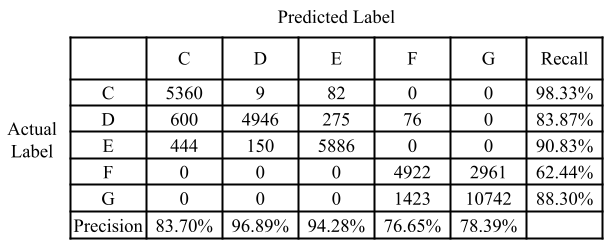}
		
	\end{center}    
\end{table}  
\begin{table}[thpb]
	\begin{center}
		\caption{Confusion Matric for the JIGSAWS Knot Tying Task with 1-second Sample Window}
		\label{tab:app_kt_conf_1s}
		\includegraphics[scale=0.33]{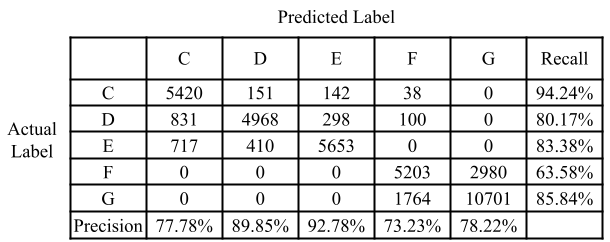}
		
	\end{center}    
\end{table}
\end{document}